\documentclass[prl,twocolumn,nofootinbib, preprintnumbers, superscriptaddress]{revtex4}

\usepackage{titlesec}
\usepackage{slashed}
\usepackage{amsmath,amssymb,slashed,braket}
\usepackage{graphicx}
\usepackage{epstopdf}
\usepackage{float,appendix}
\usepackage[colorlinks=true,
            linkcolor=blue,
            urlcolor=blue,
            citecolor=green,          
            bookmarks=true,
            bookmarksnumbered=true,
            breaklinks=true,
            pdfpagemode=Fullscreen,
            pdfstartview=FitBH]{hyperref}

\usepackage{esint}

\usepackage[normalem]{ulem}
\usepackage{color}

\definecolor{Orange}{cmyk}{0,0.61,0.87,0}
\definecolor{JungleGreen}{cmyk}{0.99,0,0.52,0}
\definecolor{OliveGreen}{cmyk}{0.64,0,0.95,0.40}
\definecolor{Brown}{cmyk}{0,0.81,1,0.60}
\definecolor{RoyalBlue}{cmyk}{0.71,0.53,0,0.12}
\definecolor{Gray}{cmyk}{0,0,0,0.40}
\definecolor{LightPink}{cmyk}{0.0,0.25,0,0}
\definecolor{LLightPink}{cmyk}{0.0,0.10,0,0}
\definecolor{LightBlue}{cmyk}{0.25,0,0,0}
\definecolor{LightGray}{cmyk}{0,0,0,0.2}

\usepackage{xcolor}
\definecolor{gesfpurple}{rgb}{0.47,0.19,0.42}

\definecolor{gesflanse}{rgb}{0.00,0.50,0.50}

\definecolor{gesfblue}{rgb}{0.08,0.42,0.76}

\definecolor{gesfred}{rgb}{1,0,0}

\definecolor{gesfwhite}{rgb}{1,1,1}

\definecolor{gesfblack}{rgb}{0,0,0}

\newcommand{\geqn}[1]{Eq.\,\hypersetup{linkcolor=blue}(\ref{#1})\hypersetup{linkcolor=blue}}
\newcommand{\gfig}[1]{{\hypersetup{linkcolor=violet}Fig.\,\ref{#1}\hypersetup{linkcolor=blue}}}

\graphicspath{{./}}

\begin{document}

\title{
Testing Electromagnetic Memory and Aharonov--Bohm Effect via Acceleration-Imprinted Superconducting Phases
}

\author{Jie Sheng}
\email{jie.sheng@ipmu.jp}
\affiliation{Tsung-Dao Lee Institute, Shanghai Jiao Tong University, Shanghai, 200240, China}
\affiliation{
Kavli IPMU (WPI), UTIAS, University of Tokyo, Kashiwa, 277-8583, Japan}

\author{Tsutomu T. Yanagida}
\email{tsutomu.tyanagida@gmail.com}
\affiliation{
Kavli IPMU (WPI), UTIAS, University of Tokyo, Kashiwa, 277-8583, Japan}
\affiliation{Tsung-Dao Lee Institute, Shanghai Jiao Tong University, Shanghai, 200240, China}

\author{Bo Gao}
\email{gaobo\underline{\,\,\,}79@sjtu.edu.cn}
\affiliation{Tsung-Dao Lee Institute, Shanghai Jiao Tong University, Shanghai, 200240, China}

\author{Hong Ding}
\affiliation{Tsung-Dao Lee Institute, Shanghai Jiao Tong University, Shanghai, 200240, China}
\affiliation{New Cornerstone Science Laboratory, Shanghai, 201201, China}

\begin{abstract}

Electromagnetic memory is an infrared observable of gauge theory associated with soft photons and large gauge transformations. Despite its fundamental theoretical importance, it has not yet been experimentally verified. From a phenomenological perspective, a transient electromagnetic configuration can leave a persistent gauge-invariant phase imprint on charged coherent states after the local field has vanished through electric Aharonov-Bohm effect. We point out that the electric field and associated gauge potential induced inside a normal conductor by gravitational acceleration can imprint this phase without requiring external phase-imprinting electric field. It can then be read out through a superconducting protocol. For representative parameters, the predicted signal can lie within the range of present sensitivities, providing a possible tabletop route toward testing both electromagnetic memory and Aharonov-Bohm effect.

\end{abstract}

\maketitle 

\section{Introduction} 
Electromagnetic memory~\cite{Kapec:2015ena,Pasterski:2015zua,Susskind:2015hpa} is a theoretically predicted effect in which an electromagnetic process leaves a persistent record after the radiative field itself has passed. In its classical form, this record appears as a permanent momentum kick or displacement of charged test particles, determined by the time integral of the radiative electric field~\cite{Bieri:2013hqa,Mao:2017axa,Sarkkinen:2018qwj}. In the language of gauge theory, the same effect can be understood as a change between the early- and late-time gauge configurations, associated with soft photons and large $U(1)$ gauge transformations~\cite{Tolish:2014bka,He:2014cra,Kapec:2014zla,Mohd:2014oja,Kapec:2015ena,Campiglia:2015qka,Pasterski:2015zua,Hawking:2016msc,Strominger:2017zoo}. Therefore, its observation is conceptually important.

The memory effect was originally proposed in the gravitational context. It predicts that the relative position of pairs of masses in space can be permanently changed by the passage of a gravitational wave~\cite{Braginsky:1985vlg,Braginsky:1987kwo,Ludvigsen:1989kg,PhysRevLett.67.1486}. Analogously, EM radiation emitted by accelerated charges can also cause a lasting change in the momentum of test particles, perpendicular to the radial direction~\cite{Bieri:2013hqa,Mao:2017axa,Sarkkinen:2018qwj}. These classical EM memory observables are, however, extremely weak and easily overwhelmed by perturbations, and therefore remain experimentally unverified~\cite{Mao:2017axa,Zosso:2025orw}.

Later research further pointed out that EM memory can also be imprinted in the quantum properties of wave functions~\cite{Susskind:2015hpa}.
The distortion of the vector potential in space by the past EM field can alter the quantum phases of the states.
It shows the quantum aspect of EM memory beyond the classical effects.
A gedanken experiment was also proposed to detect this phase difference using superconductors \cite{Susskind:2015hpa}. A related proposal suggested exploiting Josephson junction systems to detect the accumulated phase~\cite{Bachlechner:2019deb}. These proposals indicate that superconductors provide a natural quantum probe of EM memory. However, their experimental realization is difficult, because the central task is to control electromagnetic processes at small scales without introducing background potentials or switching disturbances that are indistinguishable from the memory phase.

In this paper, we propose that gravitational acceleration can provide a clean way to generate the transient electric field required for imprinting EM quantum memory. Inside a normal conductor, acceleration would induce a small but stable internal electric field due to the different responses of the ionic lattice and the electron gas. This electric field is accompanied by a gauge potential and can imprint a gauge-invariant phase difference between superconducting states placed at the two ends through the electric Aharonov--Bohm (AB) effect~\footnote{Our setup is closely related to the superconducting electric-AB proposal of Ref.~\cite{Bachlechner:2019deb}: The main difference is that the
externally charged capacitor is replaced
by an acceleration-polarized normal conductor.}. By changing the effective acceleration, for example through a free-fall sequence, the acceleration-induced electric field can be switched on and off without applying an external voltage pulse. The resulting phase shift can later be converted into a magnetic flux signal in a superconducting circuit, providing a test of both EM memory and AB effect.

Throughout the text, we use the natural unit with $c = \hbar = 1$

\section{Electromagnetic Memory Phase Imprinted in Superconductors}

According to Ginzburg–Landau theory \cite{Landau:1950lwq}, the coherent quantum state of Cooper pairs
in a superconductor can be described by the order parameter wave function 
$\psi = \sqrt{n_i} e^{i \phi_i}$, where $n_i$ is the number density of Cooper pairs and 
$\phi_i$ is its quantum phase. This macroscopic wave function effectively follow the Schrödinger equation $({\bf p}^2 / 2m_* + H_{\text{int}}) \psi = i\partial_t \psi$ where $m_* \simeq 2 m_e$ is the effective mass of a Cooper pair, ${\bf p}$ denotes its momentum, and $H_{\text{int}}$ represents the interactions present in the system \cite{FeynmanLecturesIII21,gross2016applied}. Its phase 
obeys a dynamical evolution
$\phi_i (t) = \mu_i t$, which can be defined as the dynamical phase. Here, $\mu_i$ denotes the energy eigenvalue; in the circuit context, it can also be interpreted as the electrochemical potential.

More generally, if the effect of the EM vector potential is included in the canonical momentum, the Hamiltonian becomes
$H = ({\bf p}- e^* {\bf A})^2/2m_* + H_{\text{int}}$ with $e^* \equiv 2e$ being the effective charge of a Cooper pair. Accordingly, the solution of the Schrödinger equation acquires an additional phase on top of the dynamical phase arising from the vector potential
$\psi = \sqrt{n_i} e^{i \phi_i} e^{i e^*\int^{{\bf x}_i} {\bf A}\cdot d{\bf l}}$ \cite{Sakurai:2011zz}. 
Therefore, as an observable, the gauge-invariant phase difference between two superconducting states, 
\begin{equation}
    \Delta \phi \equiv \phi_2 - \phi_1 - e^* \int_1^2  {\bf A}\cdot d{\bf l},
    \label{phasediff}
\end{equation}
receives a contribution from the line integral of the vector potential \cite{gross2016applied}.
It is consistent with the gauge-invariant two point function $\braket{\psi^\dagger (x_f) W (x_f, x_i) \psi(x_i)}$ with Wilson line $W \equiv \exp( i e^* \int_{x_i}^{x_f} A_\mu dx_\mu)$ \cite{Peskin:1995ev}.

Typically, the control of the phase difference $\Delta \phi$ in a superconducting system is achieved by applying a current or voltage modulation to a Josephson junction, through the Josephson equations $I \propto \sin \Delta \phi$ and $d\Delta \phi/dt = 2e V$ \cite{Josephson:1962zz} .
The phase can also be tuned through introducing magnetic flux in a SQUID \cite{gross2016applied}.
In these conventional settings the measured phase tracks the instantaneous voltage or magnetic flux applied to a connected circuit. The question addressed here is different: whether a transient electromagnetic configuration can leave a residual phase after the field itself has vanished.

On the other hand, if the phase difference is generated by a built-in electric field in a closed region, and the superconductors on both sides are coherent but not coupled through circuits, the phase can serve as stored memory \cite{Bachlechner:2019deb}. 
The two superconductors must be initially connected to align their phases $\Delta \phi (t_0) = 0$, thus one can perform gauge transformation to make $\phi_2 - \phi_1 = 0$, and ${\bf A} = 0$ everywhere. After that, we disconnect the superconductors and let a dynamical electric field appear and then vanish, as required by the EM memory effect \cite{Bachlechner:2019deb}. This event can be revealed as stored information in the memory phase difference of the superconductors \cite{Susskind:2015hpa}.

\begin{figure}[t]
    \centering
    \includegraphics[width=9.5cm]{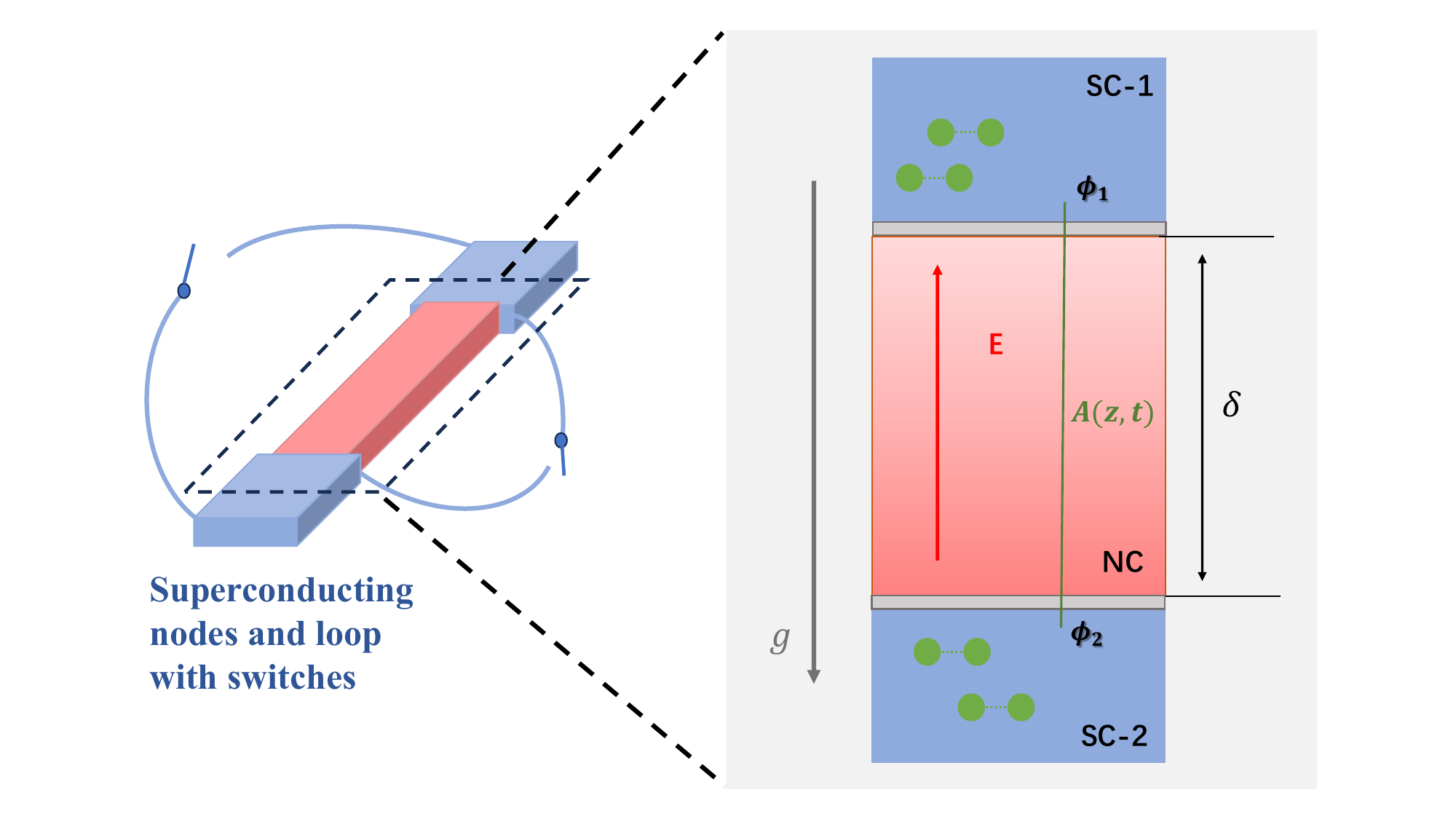}
    \caption{Cartoon of a closed electric field generating a phase difference between two superconductors through the vector potential effect. The two superconductors (blue) are connected by a superconducting loop with two switches, while the region in between is a conductor (red) and two thin insulator layers (grey).
    (The actual size of Cooper pairs is larger than that shown in the figure, and there is a high degree of overlap between them.).}
    \label{fig:gravityJJ_new}
\end{figure}

After the disconnection as shown in \gfig{fig:gravityJJ_new}, an electric field ${\bf E}$
is somehow generated and confined within the red region.
Due to the presence of ${\bf E} = - \dot{\bf A}$ with gauge choice of $A_0 = 0$, there must exist a non-trivial distribution of ${\bf A}$ inside the red
region. In addition, the magnetic field ${\bf B} = 0$ requires ${\bf A} ({\bf x}, t) = \nabla \lambda ({\bf x},t)$ inside junction region. To match the constant $|{\bf E}|$ along the $\hat z$ direction, the $\lambda$ function can be $\lambda(z,t) = - {\bf E} \cdot {\bf z}  t$. 
Outside the red region, four-vector potential is a pure gauge $A_\mu ({\bf x}) = \partial_\mu \alpha ({\bf x},t)$ since ${\bf E} = {\bf B} = 0$ in superconductors (blue region).
However, the value of $\alpha({\bf x},t)$ function needs to satisfy the continuity condition match $\lambda ({\bf x},t)$ at the boundary.

Suppose the red region in \gfig{fig:gravityJJ_new} has a length 
$\delta$, and we set $\alpha ({\bf x}_1,t)=\lambda({\bf x}_1, t)  = 0$ 
at the point ${\bf x}_1$ of superconductor $1$ adjacent to the red region. At any time $t$, at the point 
${\bf x}_2$ of superconductor $2$ adjacent to the junction, we have 
$\alpha ({\bf x}_2,t) = \lambda({\bf x}_2, t) = |{\bf E}| \delta t$ due to the continuity. 
Even if the electric field is turned off at time $\tau$, the vector potential at position ${\bf x}_2$ differs from the initial one as $\partial_\mu \alpha({\bf x}_2, \tau)$ with $\alpha({\bf x}_2, \tau) = \lambda({\bf x}_2, \tau)$. Although this pure gauge is locally unobservable, its relative action on the two previously connected superconducting condensates is retained as a gauge-invariant phase difference.
Performing a gauge transformation $A_\mu \rightarrow A_\mu - \partial_\mu \alpha$ to set $A_\mu = 0$, the field of the charged particle acquires an additional phase $\psi \rightarrow \psi e^{i e^* \alpha}$ \cite{Susskind:2015hpa}. 
The relative phase difference between the points ${\bf x}_1$ and ${\bf x}_2$ is then \cite{Susskind:2015hpa,Bachlechner:2019deb},
\begin{equation}
    \Delta \phi = e^* \alpha(x_2) = 2 e |{\bf E}| \delta t.
\label{phase_evo}
\end{equation}
This result is consistent with the gauge-invariant phase difference between the two superconductors calculated directly from \geqn{phasediff}, where the dynamical phase does not evolve, $\phi_2 = \phi_1$. Only the integral over the vector potential ${\bf A}$ contributes\footnote{We assume the electric field is confined within the red region and drops suddenly to zero at both ends. In practice, as discussed in the next section, it is rapidly screened by the conductor or superconductor near the boundaries. This does not affect the phase calculation, since ${\bf E} = -\partial_t \nabla \lambda$
and thus 
\begin{equation}
    \Delta \phi = e^*\lambda({\bf x}_2, \tau ) = e^* \int_0^\tau dt \int_{{\bf x}_1}^{{\bf x}_2} d{\bf l} \cdot {\bf E}.
\end{equation}
The dominant contribution arises from the spacial integral of electric field, while the narrow boundary region contributes negligibly.}.
Note that the phase difference evolution is numerically the same as that induced by a voltage difference $|{\bf E}| \delta$ due to the freedom of gauge choice \cite{Bachlechner:2019deb}. 
During the switching interval the phase can be computed either from the electric field integral or, equivalently, from the residual gauge transformation of the four-potential.
The memory character lies in the fact that the phase remains observable after the electric field has vanished.

In fact, this phase is generated precisely through the electric AB effect~\cite{Bachlechner:2019deb}. In the original paper \cite{PhysRev.115.485}, the electric AB effect refers to electrons feeling a potential without traversing an electric field. Since scalar and vector potentials can be related by a gauge transformation, the scenario considered here can also be seen as test of the long-standing AB effect \cite{PhysRevB.40.3491,vanOudenaarden1998,PhysRevB.67.033307,Bachlechner:2019deb}.

If an electric field lasts for a duration $\tau$, the two nodes acquire a corresponding phase difference $\Delta \phi = 2e |{\bf E}| \delta \times \tau$. 
In an ideal scenario, the memory is maintained as long as the coherence is not disturbed by the environment.
When the two superconductors are isolated, their absolute phases are undefined, and the phase difference becomes physical observable only when we re-couple the two superconductors again and measure a current or magnetic flux as explained below.

With the theory established, the central challenge is not only technical but also conceptual -- how to create a stable internal electric field within a closed region while preserving the encoded memory information. 
Because of their extreme sensitivity to EM perturbations, the electric AB phase has not been successfully demonstrated experimentally \cite{PhysRevB.40.3491,vanOudenaarden1998,PhysRevB.67.033307}. Here we propose that gravity and acceleration can act as a source of an intrinsic and stable electric field within a conducting region without introducing external electrical switching pulses.

\section{Gravity-induced Electric Field as a Memory Source}

\begin{figure}[t]
    \centering
    \includegraphics[width=8.5cm]{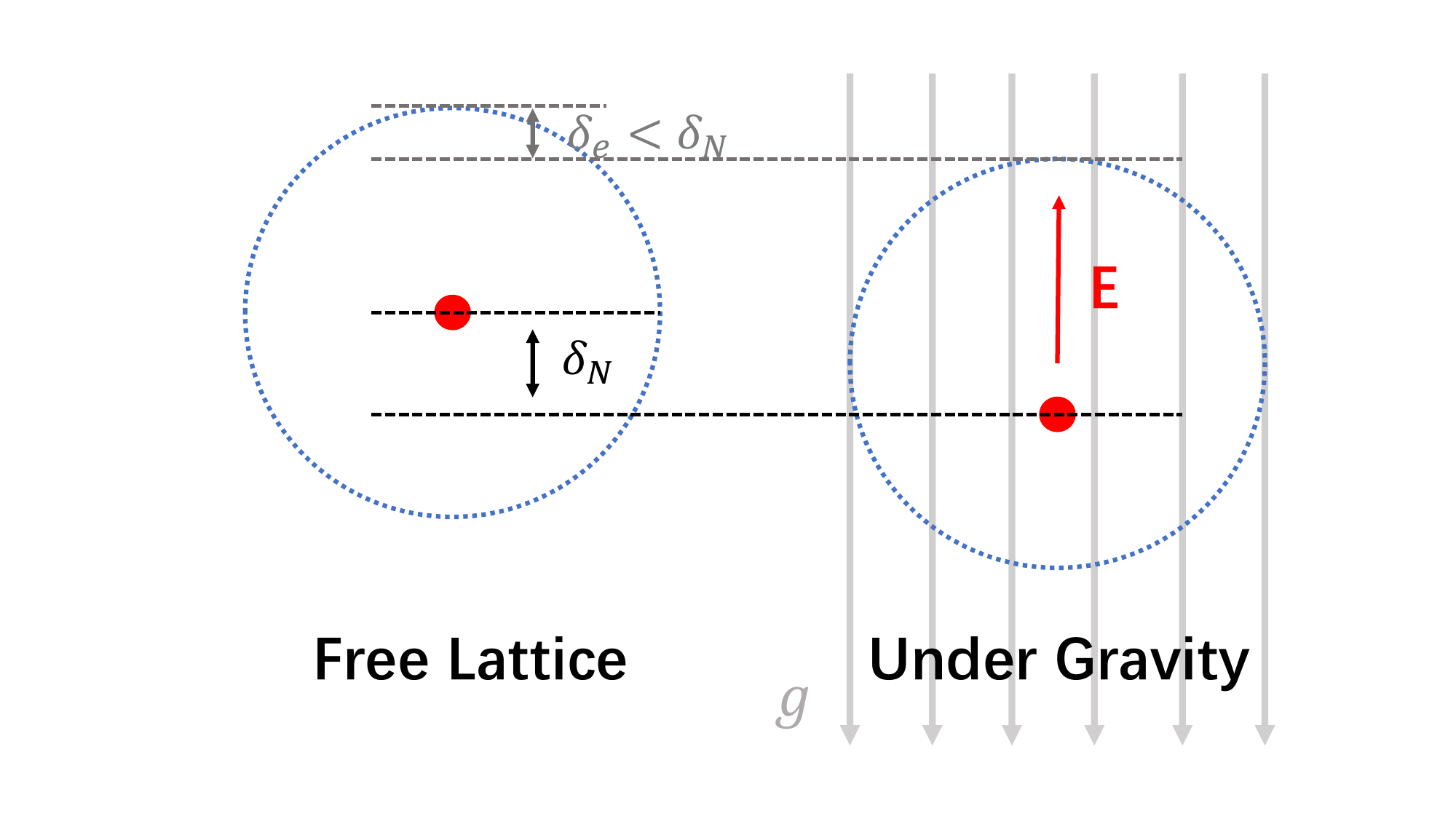}
    \caption{Microscopic mechanism of gravitationally induced electric fields in conductors \cite{SHEGELSKI2023127}.
The small red circles represent the nuclei within the atoms, while the blue dashed circles depict the electron cloud. Due to the compression of the lattice in the gravitational field, the nuclei (electrons) are displaced downward by a height $\delta_N (\delta_e)$ with $\delta_N > \delta_e$.}%
    \label{fig:field}
\end{figure}

A conductor held with proper acceleration $a_{\rm eff}$ develops a small internal electric field. In the laboratory supported on Earth, this effective acceleration originates from gravity, $a_{\rm eff} = g$, whereas in free fall $a_{\rm eff}$ vanishes.
Under Earth's gravity, atoms inside a conductor undergoes a downward compression as shown in \gfig{fig:field}, the actual density of atoms $n(z)$ along $z$-direction deviates from their equilibrium density $n_0$ in the absence of gravity, with the density increasing at lower heights $z$. The upward elastic force of solid deformation balances the gravitational force acting on its constituent atoms,
\begin{equation}
    Y \frac{\partial}{\partial z} 
    \left( \frac{n}{n_0} \right)
=
    - n_0 M g.
\end{equation}
Here, $Y$ is a function of Young's modulus and Poisson's ratio \cite{PhysRev.168.737}.

The density variation of electrons must align with the atoms to maintain electrical neutrality within the conductor. Modeled as free electron gas, 
the corresponding averaged thermal energy of the electrons in a metallic conductor is $\epsilon = 3 (3 \pi^2 n)^{2/3}/10 m_e$, and the Fermi pressure is $p = 2 n \epsilon /3$. Due to the change in density, different from nucleons, electrons will be subjected to an additional upward Fermi pressure force,
\begin{equation}
    \frac{\partial p}{\partial z}
=
    \frac{10}{9} \left( \frac{\partial n}{\partial z}\right) \epsilon
\equiv 
    \frac{10}{9} \left( \frac{\partial }{\partial z} \right) \left( \frac{n}{n_0} \right) u_e,
\end{equation}
where the energy density of free electrons is $u_e \equiv n_0 \epsilon$.
The displacements of nuclei and electrons differ under the Earth gravity, which induces an dipole-electric field ${\bf E}$ in a conductor. This electric field force acting on the electrons balances their gravitational force and the fermi pressure,
\begin{equation}
    \frac{\partial p}{\partial z}
=
    -n_0 \left( e |{\bf E}| + m_eg \right).
\end{equation}
Defining $\gamma \equiv (10/9) u_e/Y$, the above equation gives\footnote{The historical development of these discussions \cite{PhysRev.151.1067,PhysRevLett.19.1049,PhysRev.168.737,PhysRev.180.1606,PhysRevB.2.825,article_Leung,PhysRevLett.22.700,PhysRevB.1.4649} has been systematically documented in review \cite{SHEGELSKI2023127}. Here we only sketch the physical picture and conclusions.},  
\begin{equation}
    {\bf E} 
=
    \frac{g}{e} \left( \gamma M - m_e \right) \hat z.
    \label{Eg}
\end{equation}

For most metals, the parameter $\gamma$ lies in the range of $0.1 \sim 1$ \cite{PhysRev.168.737}. For the conservative benchmark $M=10\,$GeV and $\gamma=0.1$, Eq.~\eqref{Eg} gives $|{\bf E}_g|\simeq 1.1\times10^{-7}\,$V/m. The field scales linearly with both $\gamma$ and the atomic mass $M$. This effect of classical gravity in a quantum Fermi gas has received some experimental support \cite{SHEGELSKI2023127}.

On the other hand, the gravitational effect in superconductors is extremely weak. Since Cooper pairs can be treated as effective bosons free from Fermi pressure \cite{tinkham2004introduction}, the gravity-induced electric field is proportional to the electron mass and merely balances the gravitational force \cite{PhysRevB.2.825,article_Leung,JLowTempPhys.10.151,PhysRevB.48.351,rystephanick,Anandan:1985pv,Greenberger1986-GRENTA}, which is negligible compared to the nuclei-mass–enhanced field in normal conductors.

\section{Experimental Protocol and Memory Readout}

According to the equivalence principle, a uniform gravitational field is locally equivalent to a uniformly accelerated frame, up to tidal corrections. Therefore, the gravity-induced electric field discussed above should more generally be controlled by the effective proper acceleration experienced by the apparatus.
Here, we propose to use a free-fall elevator to control the proper acceleration, allowing the electric AB phase to be accumulated in a stationary supported stage and, when desired, its persistence to be tested after the source field is removed in free fall.

The relevant control parameter is the proper acceleration $a_{\rm eff}$ of the apparatus. In a supported stage $a_{\rm eff} \simeq g$, while in free fall $a_{\rm eff} \simeq 0$ up to tidal corrections. The induced field in \geqn{Eg} can be generalized to
\begin{equation}
       {\bf E}_{\rm sig} (t)
=
    \frac{a_{\rm eff} (t)}{e} \left( \gamma M - m_e \right) \hat z,
    \label{Ea} 
\end{equation}
with the time-dependent acceleration $a_{\rm eff} (t)$.
During the whole measuring process, the memory phase signal becomes, 
\begin{equation}
    \Delta \phi_{\rm sig} = e^* \delta \int dt {\bf E}_{\rm sig} (t) \cdot {\bf n}.
\end{equation}
Here, $\hat{\mathbf n}$ is the unit vector pointing from SC-1 to SC-2, such that reversing the orientation of the apparatus changes the sign of the accumulated phase.
The protocol is designed to isolate the EM memory aspect: a finite-duration electric field produces a residual gauge-invariant phase that is read out only after the field is removed. 
The role of acceleration is not to mimic
electromagnetic radiation from accelerated
charges, but to control the internal electric
field without an externally applied voltage
pulse. This does not eliminate electrical
switching of the superconducting link,
which remains necessary for phase
initialization and readout as discussed later.

The basic detecting unit in the elevator is a  superconductor-conductor system shown in \gfig{fig:gravityJJ_new} with the red region being composed of a normal conductor, and the entire device is oriented vertically with gravity. The gravity-induced electric field arises from atomic polarization rather than the motion of free charges, allowing it to be confined within the closed region and shielded on the boundary \cite{Kittel2004}. It should be noted, however, that a superconductor–conductor configuration can also form a SNS Josephson junction. According to the RSCJ model \cite{gross2016applied}, a phase difference would induce both Josephson current and displacement current, which disrupt the phase evolution. 
To avoid forming an unintended SNS Josephson weak link, the active region should be implemented as an S-I-N-I-S structure (the insulator I are shown as the gray region in \gfig{fig:gravityJJ_new}), with thin insulating barriers suppressing Cooper-pair tunneling during the imprinting stage. The normal-metal length $\delta$ controls the acceleration induced voltage drop, while the tunnel barriers suppress residual Josephson coupling. This separation should be sufficiently long as $\delta > \mathbf{O}(100)\,$nm so that during memory imprinting the superconducting nodes are exposed to the gauge potential but not phase-locked by a supercurrent.

\begin{figure}[t]
    \centering
    \hspace*{-0.6cm}
\includegraphics[width=9.5cm]{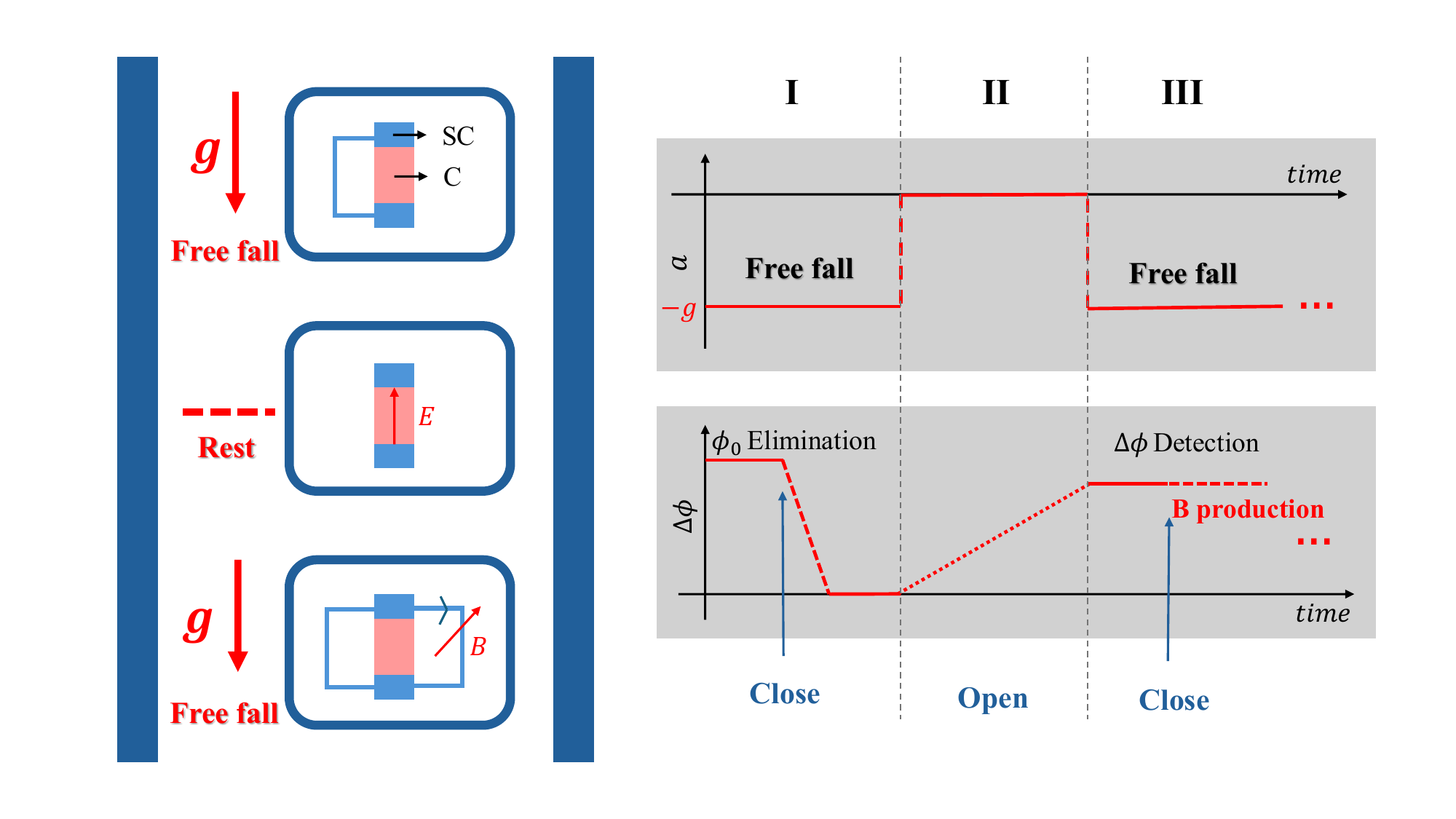}
    \caption{Cartoon of the proposed experimental scheme. The system consists of superconductor (SC, blue) and conductor (C, red) is placed inside an elevator. The idealized time evolutions of the elevator acceleration and the phase difference are shown in the right panel, without including possible background contributions.}
    \label{fig:elevator}
\end{figure}

The proposal should contain the following three steps in time sequence:
{\bf I}. There may exist a random initial phase difference $\Delta \phi_0$ between superconducting nodes, which must be eliminated. It can be achieved via connecting the nodes by superconducting wires to enable the them to form a single superconducting body with a common phase. During this procedure, the elevator is in free fall, ensuring $a_{\rm eff}=0$ that no acceleration-induced electric field arises in the conductor to induce additional phase evolution. Crucially, this circuit remains closed while the apparatus is brought from free fall to the supported stage, until the current has stabilized at zero. The landing-induced acoustic and strain transients are therefore not stored as the initial phase of the sensing interval.
{\bf II}. Afterward, the circuit is opened and the elevator keeps being rest. Once gravity acts and $a_{\rm eff}=g$, a gravity-induced electric field arises in the conductor. If the elevator remains at rest for time $\tau$. The four-vector potential $A_\mu$ keeps being distorted during this process, the gauge invariant phase difference records the memory of EM vector potential experienced by the system as $\Delta\phi_{\rm sig}=2e\delta|{\bf E}_{\rm sig}|\tau$.
{\bf III}. Finally, the elevator is set into free fall again, causing the effective gravitational force and the electric field in the conductor to vanish.
At this stage, the circuit must be reconnected after entering free fall to read out the memory phase. Thus, the stress transient from the second free-fall transition cannot be removed as in Step~I and will be addressed below~\footnote{
As an order-of-magnitude estimate, measurements of
stress-induced contact potentials~\cite{Craig1969,Leners1972} suggest
$\sim 3\,\mu\mathrm{V/atm}$.
The local gravitational-stress variation
$\Delta\sigma\sim\rho gL$, with
$\rho\sim9\times10^3\,\mathrm{kg/m^3}$ and
$L=100\,\mathrm{nm}$, then gives an additional potential
$\Delta V_{\rm CP}\sim 3\times10^{-13}\,\mathrm{V}$.
However, a contact-potential shift does not necessarily
produce an equivalent electrochemical-potential
difference between the superconducting nodes.
Parameterizing this conversion by a factor $\eta$,
the corresponding phase is estimated as
$|\Delta\phi_{\rm stress}|
\sim 0.8|\eta|
(L/100\,\mathrm{nm})
(t_{\rm eff}/1\,\mathrm{ms})$,
where $t_{\rm eff}$ is the effective duration of the
release transient.
For $|\eta|\lesssim1$, this suggests a phase of
order one radian or less, although the estimate
does not bound additional mechanical transients.
Even if a reproducible phase offset exceeds $2\pi$,
the relative phase can be extracted using standard
circular statistics and phase-unwrapping techniques
\cite{Fisher1993,Itoh1982}, provided that sufficient
phase visibility is maintained.
A detailed calculation of the device-dependent
stress coupling is left for future work and
experimental calibration.
}.

At a later time, we recouple the two superconductors so that the phase difference manifests. For example, by closing a superconducting wire loop, a supercurrent is induced, which in turn generates a magnetic flux $\Delta\Phi$ in the loop. The flux-induced phase of the loop then compensates the original phase difference to satisfy the quantization condition of the superconducting phase.
Therefore, the magnitude of the magnetic flux should satisfy $\Delta \Phi = (\Delta \phi_{\rm sig} + 2 n \pi)/2e$ \cite{FeynmanLecturesIII21,gross2016applied}.
It can be detected, for example, by using a nearby SQUID \cite{Gao:2025ryi}. 
If the elevator remains at rest for about $\tau = 1\,$ms in Step {\bf II}, the memory phase can reach 
\begin{equation}
    \Delta \phi_{\rm sig} \simeq 0.03 \times \left(\frac{\tau}{1\,\text{ms}}\right)\left(\frac{\delta}{100\,\text{nm}}\right)\left(\frac{|{\bf E}_{\rm sig}|}{10^{-7}\,\text{V/m}}\right).
    \label{frequency}
\end{equation}
This $\mathcal{O}(0.01)$ phase difference and magnetic signal can be distinguishable from the expected theoretical and experimental backgrounds.

Several device-dependent effects must be considered
when assessing the phase sensitivity.
\textit{(i) Number--phase uncertainty and charging effects.}
Usually a coherent state has
number fluctuations $\Delta N\sim\sqrt{N}$. However,
the phase uncertainty of an isolated superconducting
node also depends on its charging energy,
$E_C=e^2/(2C_\Sigma)$, where $C_\Sigma$ is the total
capacitance of the node \cite{Koch2007}.
In our protocol, the two superconducting nodes are
initially phase-locked before being isolated. Thus,
disconnection does not immediately render their
relative phase completely uncertain. Nevertheless,
the finite charging energy can cause subsequent
quantum phase spreading during the holding interval
\cite{Sols1994}. This effect can be suppressed by
increasing the effective capacitance of the nodes.
The normal-metal length $\delta$ does not
constrain the size of the superconducting nodes, which
can be implemented as much larger electrodes with
additional capacitance to suppress charging effects.

\textit{(ii) Quasiparticle poisoning.}
Another possible source of noise is quasiparticle
poisoning, which can cause discrete changes in the
charge parity of superconducting islands
\cite{Riste2013}. For stochastic poisoning events
with a total rate $\Gamma_{\rm qp}$, the probability
of at least one event during the holding interval is
\begin{equation}
P_{\rm qp}=1-e^{-\Gamma_{\rm qp}\tau}
\simeq\Gamma_{\rm qp}\tau.
\end{equation}
Therefore, one requires
$\Gamma_{\rm qp}\tau\ll1$.
Encouragingly, charge-parity switching rates below
$1\,\mathrm{Hz}$ have been experimentally achieved
in engineered superconducting circuits
\cite{Pan2022}. For a holding time of
$\tau=1\,\mathrm{ms}$, this corresponds to an event
probability below $10^{-3}$ per node.
These results demonstrate that quasiparticle
poisoning can be strongly suppressed.

Besides, any circuit system suffers from (iii) \textit{Johnson-Nyquist Noise:}
The thermal noise produces 
an inevitable background 
current, $I_T \approx e k T/ \hbar \approx 10^{-7} (T/1 K) A$. For a conservative estimation, 
we can make the low-temperature experiment setup to $T = 1\,$mK \cite{Deppner:2021fks}
and the thermal current is negligible as $I_T = 10^{-10}\,$A. In the unit of magnetic flux, state-of-the-art magnetic readout noise can be less than
$1 \mu \Phi_0/\rm{Hz}^{1/2}$ with $\Phi_0 \equiv \pi/e$~\cite{2020ApPhL.117l2601N,Malnou:2023wfo,Neidig:2025tfs}.
The benchmark phase is well above the nominal flux-resolution floor of present SQUID readout in the frequency range of $1/1\,\text{ms} = 1\,\text{kHz}$~\footnote{Usually, the changes in classical observables caused by EM memory, such as the velocity of particles, are extremely small and far below the observable level \cite{Bieri:2023btq}. Now, through the phase cumulation, a tiny effect can be converted into a measurable phase difference \cite{Cheng:2024yrn,Cheng:2024zde}.}, so the main issue is instead the rejection of backgrounds of acceleration-induced stress and readout systematics.

The primary discriminator between the signal and backgrounds is the linear dependence on the stationary holding time $\tau$. Therefore, by analyzing the time correlation, the signal can be extracted from the background.
For a fixed landing, release, and readout sequence, we can decouple the total measured phase difference $\Delta\phi_{\rm tot}$ as, 
\begin{equation}
\Delta\phi_{\rm tot}(\tau)=(\Omega_{\rm bg} \pm \Omega_g)\tau+\phi_{\rm sw}+\phi_{\rm off}+\delta\phi(\tau).
\label{eq:slope}
\end{equation}
Here, we define $\Omega_g \equiv 2e\delta |E_g|$ as the 
gravitationally induced phase-accumulation rate, while
$\Omega_{\rm bg}$ represents background contributions
that also depend linearly on the holding time $\tau$,
such as a residual voltage bias between the superconducting nodes.
Besides, $\phi_{\rm sw}$is the reproducible switching contribution, including possible electric-field backgrounds generated by mechanical stress during the acceleration-changing stage, while $\phi_{\rm off}$ is a $\tau$-independent readout offset which might be induced by mechanical vibration and acoustic phonons.
The time-dependent $\delta\phi$ contains non-reproducible residuals including the fluctuations of the previous two.

Thus, for two differential measurement with identical switching procedures and different phase evolution times $\tau_1$ and $\tau_2$,
\begin{equation}
\Delta\phi_{\rm tot}(\tau_2)-\Delta\phi_{\rm tot}(\tau_1)=\Omega_g(\tau_2-\tau_1)+\delta\phi(\tau_2)-\delta\phi(\tau_1),
\end{equation}
the repeatable release transient and readout offset can be canceled even if they are larger than the static signal.
The remaining signal and residual background can then be distinguished through repeated measurements by exploiting their different temporal correlations.
If the remaining single-shot residual has zero mean and root-mean-square(rms) width $\sigma_\phi$, averaging $N$ statistically independent repetitions gives $\tilde{\sigma}_{\phi}=\sigma_\phi/\sqrt{N}$. For example, an $\mathcal{O}(1)$ rms random background about $30$ times larger than the characteristic AB-phase variation requires $N\sim10^3$ repetitions to reach the signal scale, while $N\sim9\times10^3$ would give a nominal $3\sigma$ statistical separation. In addition to the time dependence, we can vary the conductor length and perform a similar correlation analysis.

Importantly, the linear dependence on the holding
time alone does not uniquely identify the
gravitational source $\Omega_g$ from the static parasitic voltage $\Omega_{\rm bg}$.
To distinguish these two, the protocol
must include a control measurement with the
apparatus in the opposite vertical orientation.
The gravitational contribution to the phase slope
changes sign, while orientation-independent
backgrounds remain unchanged. The gravitational
slope can therefore be extracted as
\begin{equation}
\Omega_g=\frac{\Omega_+-\Omega_-}{2},
\end{equation}
where $\Omega_\pm$ denote the measured slopes
for the two orientations.
Alternatively, a holding-time scan performed
in free fall, where $a_{\rm eff}\simeq 0$,
provides a null test of the acceleration-induced
contribution. Orientation-dependent systematic
effects must be independently characterized.

An important note is that, if the experimental target is restricted to the electric AB effect itself, the final transition to free fall is unnecessary. After landing and mechanical settling, the nodes can be phase-reset, isolated for a controlled time $\tau$, and read out while the apparatus remains in the supported stationary configuration. This eliminates the second mechanical jerk entirely. In this sense, the signal would be cleaner, and testing the AB effect could serve as a preliminary step before implementing the full protocol. This simplified AB-only protocol also requires the
aforementioned orientation-reversal measurement to
establish the gravitational origin of the observed phase.

\section{Summary and Discussions} 

Electromagnetic memory, as an infrared phenomenological prediction of field theory, is of significant theoretical interest but remains extremely difficult to probe experimentally.
Likewise, while the magnetic Aharonov--Bohm effect is a well-established quantum phenomenon, its electric counterpart has yet to receive a definitive experimental confirmation.
In this work we proposed a protocol to probe their superconducting quantum phase form. A finite-duration electric field can imprint a gauge-invariant phase difference between two previously phase-locked superconducting nodes. After the electric field is switched off, this phase remains stored in the superconducting state and can be converted into a current or flux signal upon reconnection.

The key element of the proposal is to use gravity~\footnote{Testing the universality of quantum mechanics under weak gravity remains an important subject \cite{PhysRevLett.34.1472,NESVIZHEVSKY2000754,Nesvizhevsky:2002ef,Jenke:2011zz,Abele:2012dn,Hohensee:2011yt,Chiao:2023ezj}.
The compatibility between classical gravity and quantum fermion statistics in conductors has been extensively discussed since the last century. However, this has not yet been directly verified experimentally, as it is difficult to probe the weak electric field inside a conductor \cite{PhysRevLett.21.1093,SHEGELSKI2023127}.
In this paper, we show that
these long-standing but unverified phenomena coexist and manifest within a common framework, allowing them to mutually illuminate and test each other.} and acceleration, rather than an externally applied voltage pulse, to generate the transient electric field. A normal conductor held with proper acceleration develops a small internal electric field due to the different responses of the ionic lattice and the electron gas. Switching the apparatus between a supported stage and free fall therefore provides a clean way to turn the field on and off while minimizing ordinary electromagnetic disturbances.
The dominant potential background is expected to arise from stress induced during the transition of acceleration process.

For representative parameters, the memory phase induced by AB effect can reach $\Delta \phi \sim 0.03$ on a millisecond timescale. By dfferential measurements,
this signal is above typical SQUID sensitivities and can be distinguished from backgrounds by its dependence on the holding time, effective acceleration, and length of the normal-metal region.

The existence of memory phase requires that both the phase difference between the superconductors and the vector potential in space remain coherent. However, how long such coherence can persist in space and time is ultimately a question for experiment. This experimental proposal may also suffer from acoustic phonon induced decoherence.
We take a phase accumulation time of $1\,$ms as a conservative estimate, since Josephson oscillations and the coherence of phase qubits on the millisecond scale have been extensively verified \cite{Wang:2024wjs,Watanabe:2025sxh,Hida:2025znz}.
Since repeated time-dependence measurements are required here, they can also continuously monitor whether the signal remains coherent.

\section*{Acknowledgements}

The authors thank Xin Liu, Antti Niemi, Chuan-Yang Xing very much for insightful discussions.
J. S. is supported by the Japan Society for the Promotion of Science (JSPS) as a
part of the JSPS Postdoctoral Program (Standard) with grant number: P25018, and by the World
Premier International Research Center Initiative (WPI), MEXT, Japan (Kavli IPMU).
T. T. Y. is supported by the Natural Science Foundation of China (NSFC)
under Grant No. 12175134, MEXT KAKENHI Grants No. 24H02244, and World Premier International Research Center Initiative
(WPI Initiative), MEXT, Japan. 
B. G. acknowledges support from 
Innovation Program for Quantum Science 
and Technology (No. 2021ZD0302700) and 
Cultivation Project of Shanghai 
Research Center for Quantum Sciences 
(Grant No. LZPY2024). H. D. acknowledges support from the New Cornerstone Science Foundation (No. 23H010801236).

\providecommand{\href}[2]{#2}\begingroup\raggedright\endgroup

\vspace{15mm}
\end{document}